\documentclass[aps,prl,twocolumn,showpacs]{revtex4-1}

\usepackage{bm,graphicx,textcomp,amssymb,amsmath,dcolumn,color}

\newcommand{\nn}{\nonumber\\}

\newcommand{\f}[1]{\mbox{\boldmath$#1$}}

\newcommand{\bea}{\begin{eqnarray}}
\newcommand{\ea}{\end{eqnarray}}
\newcommand{\eea}{\end{eqnarray}}

\begin{document}

\title{Dynamically assisted nuclear fusion}

\author{Friedemann Queisser and Ralf~Sch\"utzhold}

 \affiliation{Helmholtz-Zentrum Dresden-Rossendorf, 
 Bautzner Landstra{\ss}e 400, 01328 Dresden, Germany,}
 
 \affiliation{Institut f\"ur Theoretische Physik, 
 Technische Universit\"at Dresden, 01062 Dresden, Germany,} 

\affiliation{Fakult\"at f\"ur Physik,
Universit\"at Duisburg-Essen, Lotharstra{\ss}e 1, Duisburg 47057, Germany.}

\date{\today}

\begin{abstract}
We consider the prototypical deuterium-tritium fusion reaction. 
At intermediate initial kinetic energies (in the keV regime), 
a major bottle-neck of this reaction is the Coulomb barrier 
between the nuclei, which is overcome by tunneling.
Here, we study whether the tunneling probability can be 
enhanced by an additional electromagnetic field, such as 
an x-ray free electron laser (XFEL). 
We find that this dynamical assistance should be
feasible with present-day or near future technology. 
\end{abstract}


\maketitle

\paragraph{Introduction}

Tunneling is ubiquitous in physics.
Examples include field ionization in atomic physics and $\alpha$-decay 
in nuclear physics. 
The Gamov picture \cite{Gamow} explains the Geiger-Nuttall law 
\cite{Geiger+Nuttall}  
via tunneling of the $\alpha$-particle through the Coulomb barrier 
of the remaining nucleus. 
In the opposite process, nuclear fusion, the two nuclei must 
also overcome their Coulomb barrier, typically by tunneling,  
before they can fuse. 
As an extreme example, the Sauter-Schwinger effect predicts 
the creation of electron-positron pairs out of the vacuum 
by a strong electric field $E$, which can be understood as 
tunneling from the Dirac sea 
\cite{Sauter:1931,Sauter:1932,Heisenberg+Euler,Weisskopf,Schwinger}.  
The exponential dependence characteristic for tunneling leads 
to a strong suppression $P_{e^+e^-}\sim\exp\{-\pi E_S/E\}$ 
of the pair-creation probability for electric fields $E$ 
too far below the Schwinger critical field $E_S$ determined 
by the mass $m_e$ of the electron and the elementary charge 
$q$ via $E_S=m_e^2c^3/(q\hbar)\approx1.3\times10^{18}$V/m. 
Verifying this prediction has been one motivation for 
reaching these ultra-high field strengths $E$.
However, as we shall see below, the theoretical and 
experimental efforts motivated by this goal may also 
prove useful for assisting nuclear fusion. 

Even though tunneling is usually taught in the first course 
on quantum mechanics, our understanding is still far from complete, 
especially in time-dependent scenarios, 
cf.~\cite{Pimpale+Razavy,Azbel,Stovneng+Jauho,Ravazy,Kramer+Moshinsky,Kamenev}. 
Interesting phenomena in this context include 
the Franz-Keldysh effect \cite{Franz,Keldysh} 
or the B\"uttiker-Landauer traversal time \cite{Buttiker}. 
For the Sauter-Schwinger effect, it has been found that the 
pair-creation probability can be drastically enhanced by an 
additional weaker but time-dependent field 
\cite{Dynamically-assisted-Schwinger}, even if its 
frequency scale $\omega$ is well below the 
mass gap of $2m_ec^2\approx1~\rm MeV$. 
As another surprise, this enhancement mechanism, i.e.,  
the dynamically assisted Sauter-Schwinger effect, 
strongly depends on the concrete temporal (or spatio-temporal) 
dependence of the assisting field \cite{Linder},  
such as a Sauter $1/\cosh^2(\omega t)$ 
or Gaussian $\exp\{-(\omega t)^2\}$ pulse 
or a sinusoidal profile $\cos(\omega t)$.  
In the following, we study whether and how tunneling in 
nuclear fusion can be dynamically assisted, for example 
by the additional electromagnetic 
field of an x-ray free electron laser (XFEL) \cite{XFEL}.  

\paragraph{The model}

We consider deuterium-tritium fusion
\bea
{}^2_1{\rm D}+{}^3_1{\rm T}\,\to\,{}^4_2{\rm He}+{}^1_0{\rm n}+17.6~{\rm MeV}
\,,
\ea
where the initial kinetic energies $\cal E$ of the nuclei are in the keV regime 
and thus typical length scales (such as the tunneling distance) of order 
picometer.
Hence we may describe the two nuclei as non-relativistic point particles with 
masses $m_{\rm D}$ and $m_{\rm T}$ and positions $\f{r}_{\rm D}(t)$ and 
$\f{r}_{\rm T}(t)$.
Their dynamics is governed by the Lagrangian
\bea
L
&=&
\frac{m_{\rm D}}{2}\,\dot{\f{r}}_{\rm D}^2
+
\frac{m_{\rm T}}{2}\,\dot{\f{r}}_{\rm T}^2
-
V(|\f{r}_{\rm D}-\f{r}_{\rm T}|)
+
\nn
&&
+q
\dot{\f{r}}_{\rm D}\cdot\f{A}(t,\f{r}_{\rm D})
+
q\dot{\f{r}}_{\rm T}\cdot\f{A}(t,\f{r}_{\rm T})
\,,
\ea
where the potential $V(|\f{r}_{\rm D}-\f{r}_{\rm T}|)$ contains the Coulomb 
repulsion at large distances and the nuclear attraction at short distances 
(of order Fermi).
The vector potential $\f{A}$ represents the field of the XFEL.

At an initial energy $\cal E$ of 1~keV, the classical turning point 
$r_{\cal E}$ where $V(r_{\cal E})=\cal E$ (i.e., the minimum distance) 
is around 1.4~pm, which then determines the remaining tunneling distance 
(for higher energies $\cal E$, it is correspondingly smaller). 
Since the XFEL wavelength ($\geq$0.05~nm) is much larger than that, 
we may approximate the vector potential $\f{A}(t,\f{r})$ by a  
purely time-dependent field $\f{A}(t)$. 
As a result, the center of mass decouples from the relative 
coordinate $\f{r}_-=\f{r}_{\rm D}-\f{r}_{\rm T}$, 
and the dynamics of the latter is governed by 
\bea
\label{relative}
L_-
=
\frac{\mu}{2}\,\dot{\f{r}}_-^2
-
V(|\f{r}_-|)
+
q_{\rm eff}\dot{\f{r}}_-\cdot\f{A}(t)
\,,
\ea
with the reduced mass $\mu=(m_{\rm D}^{-1}+m_{\rm T}^{-1})^{-1}$ and the effective 
charge $q_{\rm eff}=q(m_{\rm T}-m_{\rm D})/(m_{\rm T}+m_{\rm D})\approx q/5$. 


\paragraph{Deformation of potential}  

Let us first estimate the tunneling probability without the $\f{A}$-field via the 
WKB approximation.
For low initial kinetic energies $\cal E$, the short-range details of the 
nuclear attraction are not important and the tunneling exponent is dominated
by the long-range behavior of $V$, which gives (for s-waves)
\bea
\label{WKB-exponent}
P\sim\exp\left\{-\pi\sqrt{\frac{2\mu c^2}{{\cal E}}}\,\alpha_{\rm QED}\right\}
\,,
\ea
where $\alpha_{\rm QED}\approx1/137$ is the fine structure constant.  
Of course, this expression is analogous to the Geiger-Nuttall law for 
$\alpha$-decay \cite{Geiger+Nuttall}.   
Inserting an energy ${\cal E}=1~\rm keV$ and the reduced mass 
$\mu\approx1.12~\rm GeV$, the above tunneling exponent is $P\sim10^{-15}$
(for ${\cal E}=10~\rm keV$, it is $10^{-5}$). 

At the classical turning point $r_{\cal E}$ (minimum distance) 
of around 1.4~pm (for an energy $\cal E$ of 1~keV), 
the Coulomb field strength is around $7\times10^{14}$~V/m.
As a result, near-future ultra-strong optical lasers or XFEL approaching 
this field strength regime can deform the potential barrier and thereby 
enhance (or suppress) tunneling significantly.
For example, for a constant electric field of $35\times10^{14}$~V/m,
the factor of $\pi$ in the exponent~\eqref{WKB-exponent} is replaced by 
$8/3\approx2.7$. 
Note that due to the exponential scaling of the tunneling probability $P$,
even moderate deformations can have a strong effect, e.g., $\pi\to8/3$ 
in the exponent~\eqref{WKB-exponent} implies $P\sim10^{-15}\to10^{-13}$.  

\paragraph{Floquet approach}

However, while the frequency of an optical laser is so low that this 
deformation can be treated within the quasi-static approximation, the temporal 
variations of an XFEL are too fast and hence should be taken into account.
In fact, as we shall see below, this time dependence can strongly enhance the 
tunneling probability.  

In order to study this enhancement, let us first assume an oscillating time
dependence $\f{A}(t)=A_z\f{e}_z\cos(\omega t)$ and use a Floquet ansatz 
(see, e.g., \cite{Grifoni,Grossmann+Haenggi}) 
\bea
\label{Floquet}
\psi(t,\f{r})=\sum\limits_{n=-\infty}^{+\infty} \psi_n(\f{r}) 
\exp\left\{-i{\cal E}t/\hbar-in\omega t\right\}
\,,
\ea
where $\f{r}=\f{r}_-$ denotes the relative coordinate from now on.
Assuming that the external vector potential $\f{A}(t)$
is a small perturbation, 
we employ perturbation theory and split the total Hamiltonian $\hat H(t)$
into the stationary unperturbed part $\hat H_0$ 
plus the time-dependent perturbation $\hat H_1(t)=\hat H_A\cos(\omega t)$. 
The zeroth order $\hat H_0\psi_0(\f{r})={\cal E}\psi_0(\f{r})$ represents 
the solution in the absence of the XFEL and we choose it to be a p-wave
$\psi_0(\f{r})=\psi_0^{\rm p}(r)\cos\vartheta$.  
Of course, for p-waves we have to take the angular momentum barrier 
into account. 
However, comparing the angular momentum barrier for $\ell=1$ with the 
Coulomb potential, we see that the latter dominates for radii larger than the 
reduced Compton wavelength $\lambdabar_{\rm C}$ divided by $\alpha_{\rm QED}$, 
in our case 24~Fermi.
Consequently, the angular momentum barrier becomes only relevant at very
short distances $\ll r_{\cal E}$. 

Following this strategy, the first Floquet side bands $\psi_{\pm1}(\f{r})$ 
are (to first order in $\f{A}$) determined by 
\bea
\label{side-bands}
\left({\cal E}-\hat H_0\pm\hbar\omega\right)\psi_{\pm1}(\f{r})
=
\hat H_A\psi_0(\f{r})
\,,
\ea
together with the appropriate boundary conditions. 
As expected from the selection rules, the first-order wave functions 
$\psi_{\pm1}(\f{r})$ contain s-wave and d-wave contributions, 
where we focus on the most important part 
$\psi_{+1}^{\rm s}(\f{r})=\psi_{+}^{\rm s}(r)$ in the following. 

Then Eq.~\eqref{side-bands} turns into an ordinary second-order 
differential equation for $\psi_{+}^{\rm s}(r)$ which can be 
solved numerically. 
However, we may also obtain an analytical estimate:
The zeroth order $\psi_0(\f{r})$ represents a wave which is 
incident with energy $\cal E$ from the outside, i.e., it is 
oscillating for radii $r$ larger than the turning point 
$r_{\cal E}$ and has an exponential (tunneling) tail for 
smaller radii $r<r_{\cal E}$.
As a result, the source term $\hat H_A\psi_0(\f{r})$ in 
Eq.~\eqref{side-bands} is negligibly small near the origin 
$r\ll r_{\cal E}$ and assumes its maximum near the turning 
point $r_{\cal E}$.

Now, let us first construct a particular solution of the 
inhomogeneous differential equation~\eqref{side-bands} 
which is also zero near the origin. 
Then, integrating equation~\eqref{side-bands} towards larger
radii, we see that this particular solution remains negligible 
until we approach the turning point $r_{\cal E}$ where the 
source term $\hat H_A\psi_0(\f{r})$ starts to play a role. 
For large radii, this particular solution then contains the 
forced oscillation with $\exp\{\pm ik_{\cal E}r\}$ corresponding
to the initial kinetic energy ${\cal E}=\hbar^2k_{\cal E}^2/(2\mu)$ 
plus the two locally homogeneous solutions with 
$\exp\{\pm ik_{{\cal E}+\hbar\omega}r\}$ corresponding to the 
higher energy 
${\cal E}+\hbar\omega=\hbar^2k_{{\cal E}+\hbar\omega}^2/(2\mu)$. 
However, this particular solution does not satisfy the correct 
boundary conditions for large radii, because we do not have an 
incident wave with this higher energy ${\cal E}+\hbar\omega$. 
Thus, in order to correct this, we have to add a homogeneous 
solution of equation~\eqref{side-bands} which precisely cancels 
this incident wave. 
This  homogeneous solution corresponds to a wave which is incident 
with energy ${\cal E}+\hbar\omega$, mostly reflected back to 
$r\to\infty$, but also contains a small tunneling amplitude at 
the origin, for which we can use the same WKB estimate as 
in~\eqref{WKB-exponent}, but now with ${\cal E}$ being replaced 
by ${\cal E}+\hbar\omega$. 

As a result, we find that the solution $\psi_{+}^{\rm s}(r)$
of equation~\eqref{side-bands} satisfying the correct boundary 
conditions must also contain a small amplitude at the origin, 
which gives us the dynamically assisted tunneling probability 
\bea
\label{Franz-Keldysh}
P\sim
\exp\left\{
-\pi\sqrt{\frac{2\mu c^2}{{\cal E}+\hbar\omega}}\,\alpha_{\rm QED}
\right\}
\,.
\ea
With an initial energy $\cal E$ of 1~keV and an XFEL frequency 
$\hbar\omega$ of 10~keV, for example, the above tunneling exponent is 
enhanced by ten orders of magnitude. 
Of course, while we are mostly interested in the exponent 
(as the leading-order contribution), one must also take the pre-factor 
in front of the exponent into account.
This pre-factor scales with $q_{\rm eff}^2A_z^2$, i.e., with the 
XFEL intensity. 
Thus the probability is proportional to the number of incident XFEL photons 
which indicates that this enhancement mechanism should also work with
incoherent photons. 

By numerically integrating equation~\eqref{side-bands}, we may arrive 
at quantitative results for the pre-factor, where we find that it actually 
grows for decreasing $\omega$, see also \cite{Ivlev+Melnikov}.  
However, if $\omega$ becomes too small, the above Floquet approach breaks 
down and it becomes necessary to consider higher bands $|n|\geq2$. 
From the lowest-order ($n=1$) result~\eqref{Franz-Keldysh} with 
${\cal E}=\omega=1~\rm keV$, for example, we conclude that the 
dynamical assistance requires electric field strengths of $10^{15}$~V/m, 
which is similar to those required for the deformation of the potential 
discussed above.
However, at those field strengths, the perturbative treatment above becomes
questionable (see the next paragraph).  

Since~\eqref{Franz-Keldysh} has the same form as~\eqref{WKB-exponent}, 
but just with an increased energy, one could be tempted to arrive at the 
simple picture that the nuclei just 
increase their initial kinetic energy by absorbing XFEL photons.
However, this simple picture can be rather misleading: 
Due to momentum conservation, the gain in kinetic energy of a nucleus by 
absorbing a keV photon is negligible.
Even if we consider the (classical) acceleration of a nucleus by an 
XFEL field consisting of many coherent photons with a frequency of 
$\hbar\omega=25~\rm keV$ and an ultra-high field strength of order 
$10^{17}$~V/m, i.e., merely a factor of ten below the Schwinger limit $E_S$,
the ponderomotive energy of the quivering motion is well below 1~keV.  

\paragraph{B\"uttiker-Landauer approach} 

In order go beyond the lowest-order Floquet approach above,  
we study the WKB exponent $S(t,\f{r})$ in a space-time dependent setting. 
Considering a central collision of the two nuclei along the $z$-axis, 
we assume vanishing angular momentum, i.e., 
$\partial_\vartheta S=\partial_\varphi S=0$. 
However, we have checked that including an angular dependence 
such as $S=S(t,r,\vartheta)$ does not affect the following results 
significantly -- which is consistent with our previous observation 
that the angular momentum barrier is not crucial for the parameters 
considered here. 

Employing the WKB ansatz $\psi={\cal A}\exp\{iS/\hbar\}$, 
we obtain the usual eikonal (Hamilton-Jacobi) equation 
\bea
\label{Hamilton-Jacobi}
\partial_t S(t,r)+\frac{[\partial_rS(t,r)-q_{\rm eff}A_z(t)]^2}{2\mu}+V(r)=0
\,,
\ea
with the static potential barrier $V(r)$ while the time-dependent 
XFEL field is represented by $A_z(t)$, cf.~\cite{Schneider}. 
As the next step (see also \cite{Buttiker,Fisher,Ivlev,Tanizawa}), 
we split the eikonal function 
$S(t,r)=S_0(t,r)+S_1(t,r)$ into the zeroth-order solution 
$S_0(t,r)$ of the static tunneling problem 
\bea
\partial_t S_0+\frac{(\partial_rS_0)^2}{2\mu}+V(r)=0
\,,
\ea
with $\partial_t S_0=-{\cal E}$, plus the corrections $S_1(t,r)$
induced by the XFEL field $A(t)$. 
Linearizing~\eqref{Hamilton-Jacobi}
in those quantities $S_1$ and $A$ yields the first-order equation  
\bea
\left(
\frac{\partial}{\partial t}+\frac{\partial_rS_0}{\mu}
\frac{\partial}{\partial r}
\right)S_1(t,r)
=
q_{\rm eff} A_z(t)\frac{\partial_rS_0}{\mu}
\,.
\ea
Employing the boundary condition $S_1(t,r_{\cal E})=0$, 
this equation has the solution 
\bea
\label{S_1}
S_1(t,r)=
q_{\rm eff}\int\limits_{r_{\cal E}}^{r} dr' 
A_z\left[t-\tau(r)+\tau(r')\right]
\,, 
\ea
with the well-known WKB expression \cite{Buttiker,Kira,Ganichev} 
\bea
\tau(r)=\int\limits_{r_{\cal E}}^{r} 
\frac{dr'}{\sqrt{2[{\cal E}-V(r')]/\mu}} 
\,\leadsto\,
\frac{d\tau}{dr}=\frac{\mu}{\partial_rS_0}
\,.
\ea
For classically allowed propagation ${\cal E}>V$, all the quantities 
$S_0(r)$ and $\tau(r)$ and thus also $S_1(t,r)$ are real.
For tunneling ${\cal E}<V$, however, $S_0(r)$ and $\tau(r)$ become 
imaginary and thus $S_1(t,r)$ will be complex in general.
Very analogous to the Sauter-Schwinger effect, the imaginary part of 
$S_1(t,r)$ then determines the enhancement (or suppression) of the 
tunneling probability. 
Note that $\tau$ is precisely the B\"uttiker-Landauer traversal time for 
tunneling, i.e., the imaginary turning time in the instanton picture. 

According to Eq.~\eqref{S_1}, the tunneling exponent is determined 
by the analytic continuation of the vector potential $A(t)$ to complex times 
(see also \cite{Palomares-Baez}), again in close analogy to the Sauter-Schwinger effect. 
As a result, we also find a qualitative difference \cite{Linder} between a 
Sauter $E(t)=\dot A(t)=E_0/\cosh^2(\omega t)$ and a Gaussian pulse 
$E(t)=E_0\exp\{-(\omega t)^2\}$ as well as a sinusoidal profile 
$E(t)=E_0\cos(\omega t)$ here. 
Let us first consider a sinusoidal profile which grows exponentially 
as $\exp\{\omega|\tau|\}$ for large imaginary times $\tau$. 
In analogy to Eq.~\eqref{WKB-exponent}, we may estimate the maximum 
imaginary turning time (again neglecting the finite size of the nuclei) 
via 
\bea 
\frac{{\cal E}|\tau|}{\hbar}
=
\frac{\pi}{4}\sqrt{\frac{2\mu c^2}{{\cal E}}}\,\alpha_{\rm QED}
\,.
\ea
Apart from the factor 1/4, we find the same expression as in the WKB 
tunneling exponent~\eqref{WKB-exponent}.
For ${\cal E}=1~\rm keV$, we get ${\cal E}|\tau|/\hbar\approx8.6$. 
Thus, for frequencies $\omega$ in the keV regime, $\omega|\tau|$
is a large number, which allows us to approximate our 
result~\eqref{S_1} further.
Calculating $S_1$ near the origin, the integral~\eqref{S_1} 
receives its maximum contribution near the turning point $r_{\cal E}$
(similar to the Floquet approach above). 
For an oscillating time-dependence $A_z(t)$, we may thus estimate 
this integral by
\bea
\label{large-omega}
\frac{S_1}{\hbar}\approx
\frac{iq_{\rm eff}A_ze^{i\omega t}{\cal E}^2}{2\mu c\alpha_{\rm QED}(\hbar\omega)^2}
\exp\left\{
\hbar\omega\,\frac{\pi}{4}\sqrt{\frac{2\mu c^2}{{\cal E}^3}}\,\alpha_{\rm QED}
\right\}.
\ea
Apart from the WKB pre-factor $\cal A$, the time-average of the probability 
$|\exp\{iS_0/\hbar+iS_1/\hbar\}|^2$ is given by the zeroth-order term 
$\exp\{-2|S_0|\}$ multiplied by ${\mathfrak I}_0(2|S_1|/\hbar)$, 
where ${\mathfrak I}_0$ is the modified Bessel function of the first kind. 
For small arguments, it behaves as $1+|S_1|^2/\hbar^2$ and for large arguments, 
it scales with $\exp\{2|S_1|/\hbar\}/\sqrt{4\pi|S_1|/\hbar}$. 
Note, however, that our linearized approach~\eqref{S_1} breaks down 
when $|S_1|$ becomes too large. 
The double exponential dependence of the probability on $\omega$ 
is typical for the B\"uttiker-Landauer approach (in oscillating fields) 
and shows that the required field strength is actually weaker than 
expected from the lowest-order Floquet approach above. 

The dynamical assistance sets in when $S_1/\hbar$ approaches order unity.
For ${\cal E}=\omega=1~\rm keV$, this requires fields strengths of order 
$10^{13}~\rm V/m$.
For higher frequencies $\omega$, the necessary field strengths are even lower, 
e.g., for $\omega=2~\rm keV$, we have $10^{10}~\rm V/m$.
Turning the argument around, we find that the threshold frequency $\omega_*$ 
where the enhancement mechanism sets in is determined by the inverse 
B\"uttiker-Landauer traversal time $1/|\tau|$ multiplied by $\ln E_0$. 
%
This is very reminiscent of the dynamically assisted Sauter-Schwinger effect 
for an oscillatory time dependence \cite{Linder}. 
Indeed, we find the same qualitative dependence on the pulse shape in both cases: 
For a Gaussian profile $E(t)=E_0\exp\{-(\omega t)^2\}$, 
the threshold frequency $\omega_*$ scales with $\omega_*\sim\sqrt{\ln E_0}/|\tau|$, 
while $\omega_*\sim1/|\tau|$ is nearly independent of the field strength $E_0$ 
for a Sauter pulse $E(t)=E_0/\cosh^2(\omega t)$.

\paragraph{Assistance by electrons} 

For an XFEL, time-dependences such as a Gaussian or Sauter pulse 
may be hard to realize experimentally. 
However, the Coulomb field of a particle such as an electron 
passing through (or close by) the smallest gap of the two nuclei would more 
correspond to a pulse-like time-dependence. 
Of course, the assumption of an external (i.e., classical) and 
spatially homogeneous field describes an XFEL field quite well, 
but is not such a good approximation for the Coulomb field of an electron.

Nevertheless, one would expect that the dynamical assistance mechanism 
does also apply (qualitatively) to this case, 
see also~\cite{Ivlev+Gudkov}. 
In order to obtain a first rough estimate, let us employ time-dependent 
perturbation theory with respect to the Coulomb interaction between 
the electrons and the nuclei.
The $\hat H_0$-problem of the two nuclei could in principle again be 
diagonalized in terms of center-of-mass and relative coordinates.
However, let us simplify this problem even more by fixing the position of the 
tritium nucleus (formally corresponding to the limit $m_{\rm T}\to\infty$)
and considering the motion of the deuterium nucleus in the external 
potential $V(\f{r}_{\rm D})$.
In second quantization, the Coulomb interaction Hamiltonian reads 
\bea
\label{Coulomb}
\hat H_{\rm int}=-q^2\int d^3r_{\rm D}\int d^3r_e\,
\frac{\hat\varrho_{\rm D}(\f{r}_{\rm D})\hat\varrho_e(\f{r}_e)}
{4\pi\varepsilon_0|\f{r}_{\rm D}-\f{r}_e|}
\,,
\ea
where $\hat\varrho_{\rm D}(\f{r}_{\rm D})=
\hat\Psi^\dagger_{\rm D}(\f{r}_{\rm D})\hat\Psi_{\rm D}(\f{r}_{\rm D})$
is the deuterium and 
$\hat\varrho_e(\f{r}_e)=\hat\Psi^\dagger_e(\f{r}_e)\hat\Psi_e(\f{r}_e)$
the electron density operator.
Let us consider the transition from an initial electron state with the 
energy ${\cal E}_e^{\rm in}$ to a final state with the energy 
${\cal E}_e^{\rm out}={\cal E}_e^{\rm in}-\Delta{\cal E}$. 
Then, the excess energy $\Delta{\cal E}$ is transferred to the deuterium. 
Its initial state is incident with an initial energy $\cal E$.
As before, the associated wave function decays exponentially for 
$|\f{r}_{\rm D}|<r_{\cal E}$. 
As the final state, we consider a wave function which is peaked near 
the origin (due to the nuclear attraction by the tritium) and decays 
exponentially for larger radii (inside the Coulomb barrier).
However, due to the excess energy $\Delta{\cal E}$, this final state 
has an energy ${\cal E}+\Delta{\cal E}$ and thus its exponential 
decay is slower and given by~\eqref{Franz-Keldysh} with $\hbar\omega$ 
being replaced by $\Delta{\cal E}$.
Hence, the spatial overlap integral over $r_{\rm D}$ is again peaked near 
the turning point $|\f{r}_{\rm D}|\approx r_{\cal E}$ and yields an 
exponential suppression as in~\eqref{Franz-Keldysh}.
The remaining $r_e$-integral is not exponentially suppressed and 
mainly determined by the probability that the electron is indeed 
close enough to assist dynamically. 
In this case, the field strength of the electron is also large enough. 

\paragraph{Conclusions and outlook} 

Even though nuclear physics is customarily associated with very high 
field strengths and energies (in the MeV to GeV range), we find that 
nuclear fusion can be assisted at much lower scales, which should 
become available with present-day or near future XFEL facilities 
(or with electrons), cf.~\cite{Mimura}.  
Apart from the deformation of the potential barrier, the 
time-dependence plays a crucial role for assisting tunneling 
through the Coulomb barrier -- in close analogy to the dynamically 
assisted Sauter-Schwinger effect. 

Within the lowest-order Floquet approximation, we found that the 
tunneling exponent is enhanced according to~\eqref{Franz-Keldysh}
where the replacement ${\cal E}\to{\cal E}+\hbar\omega$ 
is typical for the Franz-Keldysh effect to lowest order, 
which describes dynamically assisted tunneling in the 
perturbative regime. 
For higher orders, one would expect terms with 
${\cal E}\to{\cal E}+2\hbar\omega$ and so on, where the exponential 
enhancement is even stronger while the pre-factor is also stronger 
suppressed (e.g., with $q_{\rm eff}^4\f{A}^4$) for low intensities. 
As in the dynamically assisted Sauter-Schwinger effect, one would expect that 
higher orders can dominate in this case, cf.~\cite{Torgrimsson}.  
In order to go beyond the lowest-order Floquet approximation, we generalized the 
B\"uttiker-Landauer approach to this case and derived the first corrections 
$S_1$ to the tunneling exponent in~\eqref{S_1}.
%

The proposed dynamical assistance mechanism should also work for other 
fusion reactions. 
An important example is deuterium-deuterium fusion. 
In this case, the above approximation $\f{A}(t,\f{r})\approx\f{A}(t)$ 
is not adequate because $q_{\rm eff}=0$ and we have to include the 
spatial dependence of the XFEL field.
For an XFEL wavelength of 50~pm and distances of order 1~pm, this 
results in a suppression by a factor of around 1/50, which is partly 
compensated by the fact that $q_{\rm eff}\approx q/5$ is now replaced 
by $q$.
On the other hand, this suppression does not apply to the dynamical 
assistance by electrons sketched in~\eqref{Coulomb}. 

In summary, our understanding of tunneling is still far from complete 
and offers surprises which motivate further studies. 
For example, the limitation of perturbative and linearized approaches 
necessitate the development of fully non-perturbative methods, 
perhaps in analogy to the world-line instanton technique in the 
Sauter-Schwinger effect, see, e.g., 
\cite{Dunne+Schubert,Dunne+Wang+Schubert,Kim+Page,Ilderton1,Ilderton2,
Keski-Vakkuri+Kraus}. 

\paragraph{Acknowledgements}  

The authors acknowledge fruitful discussions with R.~Sauerbrey and 
financial support by the German Research Foundation DFG 
(grants 278162697 -- SFB 1242, 398912239).

\end{document}